\documentclass[12pt,letterpaper]{article}
\usepackage{epsfig,rotating,setspace,latexsym,amsmath,epsf,amssymb,amsfonts,bm,theorem,cite,caption,subcaption,enumerate,longtable,accents}
\usepackage{algorithm,algorithmic,graphicx,epsf,authblk,epstopdf,url,color,multirow}

\newtheorem{theorem}{Theorem}

\newtheorem{remark}{Remark}
\newtheorem{lemma}{Lemma}
\newenvironment{Proof}[1]{\medskip\par\noindent{\bf Proof:\,}\,#1}{{\mbox{\,$\blacksquare$}\par}}

\newcommand{\cs}{{\mathcal{S}}}

\newcommand{\st}{{\text{s.t.}}}

\allowdisplaybreaks

\begin{document}

\title{The Capacity of Private Information Retrieval from Heterogeneous Uncoded Caching Databases\thanks{This work was supported by NSF Grants CNS 15-26608, CCF 17-13977 and ECCS 18-07348. A shorter version is submitted to IEEE ISIT 2019.}}

\author{Karim Banawan \quad Batuhan Arasli \quad Yi-Peng Wei \quad Sennur Ulukus\\
	\normalsize Department of Electrical and Computer Engineering\\
	\normalsize University of Maryland, College Park, MD 20742 \\
	\normalsize {\it kbanawan@umd.edu ~~ barasli@umd.edu ~~ ypwei@umd.edu ~~ ulukus@umd.edu}}

\maketitle

\begin{abstract}
We consider private information retrieval (PIR) of a single file out of $K$ files from $N$ non-colluding databases with \emph{heterogeneous storage constraints} $\bm{m}=(m_1, \cdots, m_N)$. The aim of this work is to jointly design the content placement phase and the information retrieval phase in order to minimize the download cost in the PIR phase. We characterize the optimal PIR download cost as a linear program. By analyzing the structure of the optimal solution of this linear program, we show that, surprisingly, the optimal download cost in our heterogeneous case matches its homogeneous counterpart where all databases have the same average storage constraint $\mu=\frac{1}{N} \sum_{n=1}^{N} m_n$. Thus, we show that there is no loss in the PIR capacity due to heterogeneity of storage spaces of the databases. We provide the optimum content placement explicitly for $N=3$.
\end{abstract}

\section{Introduction}
The problem of private information retrieval (PIR), introduced in \cite{ChorPIR}, has attracted much interest in the information theory community with leading efforts \cite{RamchandranPIR, YamamotoPIR, VardyConf2015, RazanPIR,JafarPIRBlind}. In the classical setting of PIR, a user wants to retrieve a file out of $K$ files from $N$ databases, each storing the same content of entire $K$ files, such that no individual database can identify the identity of the desired file. Sun and Jafar \cite{JafarPIR} characterized the optimal normalized download cost of the classical setting to be $D^*=1+\frac{1}{N}+\cdots+\frac{1}{N^{K-1}}$. Fundamental limits of many interesting variants of the PIR problem have been investigated in \cite{JafarColluding, arbitraryCollusion, RobustPIR_Razane, symmetricPIR, KarimCoded,arbmsgPIR, MultiroundPIR, codedsymmetric, wang2017linear, codedcolluded, MPIRjournal, BPIRjournal, symmetricByzantine, codedcolludingZhang, MPIRcodedcolludingZhang, CodeColludeByzantinePIR, tandon2017capacity,KimCache, wei2017fundamental, kadhe2017private, chen2017capacity, wei2017capacity, wei2017fundamental_partial, StorageConstrainedPIR_Wei, SI_Gastpar, mirmohseni2017private, PrivateSearch, abdul2017private, StorageConstrainedPIR, KarimAsymmetricPIR, PIR_WTC_II, noisyPIR, SecurePIR, securePIRcapacity, securestoragePIR, XSTPIR, Tian_upload, Staircase_PIR, PIR_cache_edge, Kumar_PIRarbCoded, PIR_decentralized, TamoISIT, LiConverse,  PIR_lifting, PIR_networks, Karim_nonreplicated}.

A common assumption in most of these works is that the databases have sufficiently large storage space that can accommodate all $K$ files in a \emph{replicated} manner. This may not be the case for peer-to-peer (P2P) and device-to-device (D2D) networks, where information retrieval takes place directly between the users. Here, the user devices (databases) will have \emph{limited} and \emph{heterogeneous} sizes. This motivates the investigation of PIR from databases with \emph{heterogeneous storage constraints}. In this work, we aim to jointly design the storage mechanism (content placement) and the information retrieval scheme such that the normalized PIR download cost is minimized in the retrieval phase.

Reference \cite{StorageConstrainedPIR} studies PIR from \emph{homogeneous storage-limited} databases. In \cite{StorageConstrainedPIR}, each database has the \emph{same} limited storage space of $\mu KL$ bits with $0 \leq \mu \leq 1$, where $L$ is the message size (note, perfect replication would have required $\mu=1$). The goal of \cite{StorageConstrainedPIR} is to find the optimal centralized uncoded caching scheme (content placement) that minimizes the PIR download cost. \cite{StorageConstrainedPIR} shows that symmetric batch caching scheme of \cite{Caching_Maddah_Ali} for content placement together with Sun-Jafar scheme in \cite{JafarPIR} for information retrieval result in the lowest normalized download cost. \cite{StorageConstrainedPIR} characterizes the optimal storage-download cost trade-off as the lower convex hull of $N$ pairs $(\frac{t}{N}, 1+\frac{1}{t}+\cdots+\frac{1}{t^{K-1}})$, $t=1,\cdots,N$.

Meanwhile, the content assignment problem for \emph{heterogeneous} databases (caches) is investigated in the context of coded caching in \cite{Yener_heterogeneous}. In the coded caching problem \cite{Caching_Maddah_Ali}, the aim is to jointly design the placement and delivery phases in order to minimize the traffic load in the delivery phase during peak hours. Reference \cite{Yener_heterogeneous} proposes an optimization framework where placement and delivery schemes are optimized by solving a linear program. Using this optimization framework, \cite{Yener_heterogeneous} investigates the effects of heterogeneity in cache sizes on the delivery load memory trade-off with uncoded placement.

\begin{figure}[t]
	\centering
	\includegraphics[width=0.75\textwidth]{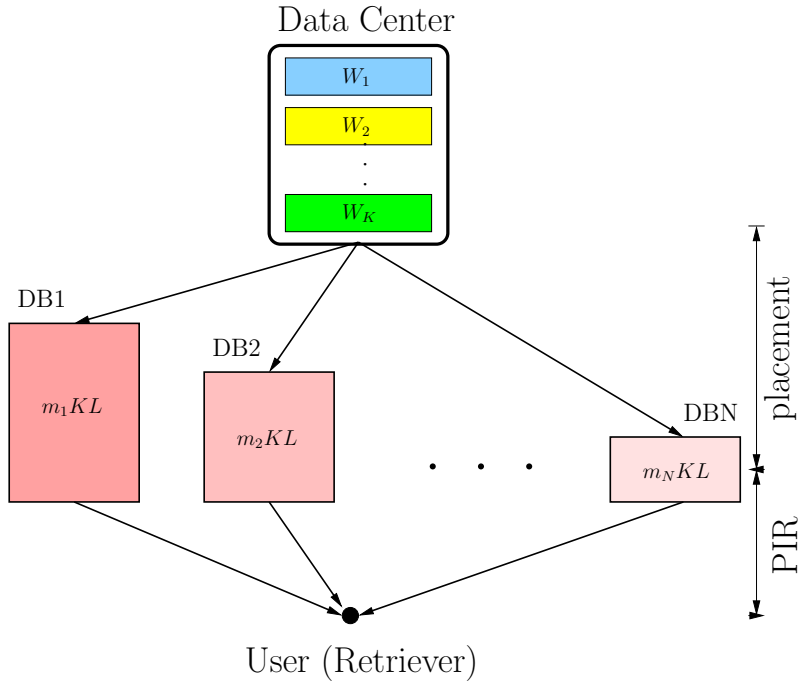}
	\caption{PIR from databases with heterogeneous storage sizes.}
	\label{hetero}
\end{figure}

In this paper, we investigate PIR from databases with heterogeneous storage sizes (see Fig.~\ref{hetero}). The $n$th database can accommodate $m_n K L$ bits, i.e., the storage system is constrained by the storage size vector $\bm{m}=(m_1, \cdots, m_N)$. We aim to characterize the optimal normalized PIR download cost of this problem, and the corresponding optimal placement and optimal retrieval schemes. We focus on uncoded placement as in \cite{StorageConstrainedPIR} and \cite{Yener_heterogeneous}.

Motivated by \cite{Yener_heterogeneous}, we first show that the optimal normalized download cost is characterized by a linear program. For the achievability, each message is partitioned into $2^N-1$ partitions (the size of the power set of $[N]$, denoted $\mathcal{P}([N])$). For every partition, we apply the Sun-Jafar scheme \cite{JafarPIR}. The linear program arises as a consequence of optimizing the achievable download cost with respect to the partition sizes subject to the storage constraints. For the converse, we slightly modify the converse in \cite{StorageConstrainedPIR} to be valid for the heterogeneous case. These achievability and converse proofs result in exactly the same linear program, yielding the \emph{exact capacity} for this PIR problem for all $K$, $N$, $\bm{m}$. Interestingly, this is unlike the caching problem in \cite{Yener_heterogeneous} with no privacy requirements, where the linear program is only an achievability, and is shown to be the exact capacity only in special cases.

By studying the properties of the solution of the linear program, we show that, surprisingly, the optimal normalized download cost for the heterogeneous problem is identical to the optimal normalized download cost for the corresponding homogeneous problem, where the homogeneous storage constraint is $\mu=\frac{1}{N} \sum_{n=1}^{N} m_n$ for all databases. This implies that there is no loss in the PIR capacity due to heterogeneity of storage spaces of the databases. In fact, the PIR capacity depends only on the sum of the storage spaces and does not depend on how the storage spaces are distributed among the databases. The general proof for this intriguing result is a consequence of an existence proof for a positive linear combination using the theory of positive linear dependence in \cite{Davis_PosLin} (and using Farkas' lemma \cite{boyd2004convex} as a special case) for the constraint set of the linear program. As a byproduct of the structural results, we show that, for the optimal content assignment, at most two consecutive types of message partitioning exist, i.e., message $W_k$ should be partitioned such that there are repeated partitions over $i$ databases and at most one more repeated partitions over $i+1$ databases for some $i$, where $i \in \{1,\cdots,N\}$. While for general $N$ we show the existence of an optimal content placement that attains the homogeneous PIR capacity, for $N=3$, we provide an explicit (parametric in $\bm{m}$) optimal content placement.

\section{System Model}
We consider PIR from databases with heterogeneous sizes; see Fig.~\ref{hetero}. We consider a storage system with $K$ i.i.d.~messages (files). The $k$th message is of length $L$ bits, i.e.,
\begin{align}
H(W_1, \cdots, W_K)=KL, \quad H(W_k)=L, \quad k \in [K]
\end{align}

The storage system consists of $N$ non-colluding databases. The storage size of the $n$th database is limited to $m_n KL$ bits, for some $0 \leq m_n \leq 1$. Specifically, we denote the contents of the $n$th database by $Z_n$, such that,
\begin{align}
H(Z_n) \leq m_n KL, \quad n \in [N]
\end{align}

The system operates in two phases: In the placement phase, the data center (content generator) stores the message set in the $N$ databases, in such a way to minimize the download cost in the retrieval phase subject to the heterogeneous storage constraints. The placement is done in a \emph{centralized} fashion \cite{Caching_Maddah_Ali}. The user (retriever) has no access to the data center. Here, we focus on uncoded placement as in \cite{StorageConstrainedPIR, Yener_heterogeneous}, i.e., file $W_k$ can be partitioned as,
\begin{align}\label{uncoded}
W_k=\bigcup_{\cs \subseteq [N]} W_{k,\cs}
\end{align}
where $W_{k,\cs}$ is the set of $W_k$ bits that appear in the database set $\cs \subseteq \mathcal{P}([N])$, where $\mathcal{P}(\cdot)$ is the power set. $H(W_{k,\cs})=|W_{k,\cs}|L$, where $0 \leq |W_{k,\cs}| \leq 1$. Under an uncoded placement, we have the following message size constraint,
\begin{align}\label{msg_size}
1=\frac{1}{KL}\sum_{k=1}^K H(W_k)=\frac{1}{KL} \sum_{k=1}^{K} \sum_{\cs \subseteq [N]} H(W_{k,\cs})=\sum_{\cs \subseteq [N]} \alpha_\cs
\end{align}
where $\alpha_\cs=\frac{1}{K} \sum_{k=1}^{K} |W_{k,\cs}|$. In addition, we have the individual database storage constraints,
\begin{align}\label{individual}
m_n \geq \frac{1}{KL} H(Z_n)=\sum_{\substack{\cs \subseteq [N], n \in \cs}} \alpha_\cs, \quad n \in [N]
\end{align}

In the retrieval phase, the user is interested in retrieving $W_\theta$, $\theta \in [K]$ privately. The user submits a query $Q_n^{[\theta]}$ to the $n$th database. Since the user has no information about the files, the messages and queries are statistically independent, i.e.,
\begin{align}
	I(W_{1:K};Q_{1:N}^{[\theta]})=0
\end{align}
The $n$th database responds with an answer string, which is a function of the received query and the stored content, i.e.,
\begin{align}
H(A_n^{[\theta]}|Q_n^{[\theta]},Z_n)=0, \quad n \in [N]
\end{align}

To ensure privacy, the query submitted to the $n$th database when intended to retrieve $W_\theta$ should be statistically indistinguishable from the one when intended to retrieve $W_{\theta'}$, i.e.,
\begin{align}\label{privacy}
(Q_n^{[\theta]},A_n^{[\theta]},W_{1:K}) \sim (Q_n^{[\theta']},A_n^{[\theta']},W_{1:K}), \quad \theta,\theta' \in [K]
\end{align}
where $\sim$ denotes statistical equivalence.

The user needs to decode the desired message $W_\theta$ reliably from the received answer strings, consequently,
\begin{align}\label{reliability}
	H(W_\theta|Q_{1:N}^{[\theta]},A_{1:N}^{[\theta]})=o(L)
\end{align}
where $\frac{o(L)}{L} \rightarrow 0$ as $L \rightarrow \infty$.

An achievable PIR scheme satisfies constraints \eqref{privacy} and \eqref{reliability} for some file size $L$. The download cost $D$ is the size of the total downloaded bits from all databases,
\begin{align}
D=\sum_{n=1}^{N} H(A_n^{[\theta]})
\end{align}

For a given storage constraint vector $\bm{m}$, we aim to jointly design the placement phase (i.e., $Z_n$, $n \in [N]$) and the retrieval scheme to minimize the normalized download cost $D^*=\frac{D}{L}$ in the retrieval phase.

\section{Main Results}
Theorem~\ref{Thm1} characterizes the optimal download cost under heterogeneous storage constraints in terms of a linear program. The main ingredients of the proof of Theorem~\ref{Thm1} are introduced in Section~\ref{capacity} for $N=3$, and the complete proof is given in Section~\ref{general-capacity} for general $N$.

\begin{theorem}\label{Thm1}
	For PIR from databases with heterogeneous storage sizes $\bm{m}=(m_1,\cdots,m_N)$, the optimal normalized download cost is the solution of the following linear program,
	\begin{align}\label{general_achievable}
	\min_{\alpha_\cs \geq 0} \quad &\sum_{\ell=1}^{N} \sum_{\cs: |\cs|=\ell} \alpha_\cs \left(1+\frac{1}{\ell}+\cdots+\frac{1}{\ell^{K-1}}\right)\notag\\
	\st  \quad &\sum_{\cs:|\cs| \geq 1} \alpha_\cs=1 \notag\\
	 \quad &\sum_{\cs: n \in \cs} \alpha_\cs \leq m_n, \quad  n \in [N]
	\end{align}
	where $\cs \in \mathcal{P}([N])$.
\end{theorem}

Theorem~\ref{Thm2} shows the equivalence between the optimum download costs of the heterogeneous and homogeneous problems. The proof of Theorem~\ref{Thm2} is given in Section~\ref{equivalence}.

\begin{theorem}\label{Thm2}
    The normalized download cost of the PIR problem with heterogeneous storage sizes $\bm{m}=(m_1, \cdots, m_N)$ is equal to the normalized download cost of the PIR problem with homogeneous storage sizes $\mu=\frac{1}{N} \sum_{n=1}^{N} m_n$ for all databases, i.e., $D^*(\bm{m})=D^*(\bar{\bm{m}})$, where $\bar{\bm{m}}$ is such that $\bar{m}_n=\mu$, for $n=1,\cdots,N$.
\end{theorem}

\begin{remark}
Theorem~\ref{Thm2} implies that the storage size asymmetry does not hurt the PIR capacity, so long as the placement phase is optimized. This is unlike, for instance, access asymmetry in the case of replicated databases \cite{KarimAsymmetricPIR}. This is also unlike, as another instance, non-optimized content placement even for symmetric database sizes \cite{Karim_nonreplicated}.
\end{remark}

\begin{remark}
Stronger than what is stated, i.e., the equivalence between heterogeneous and homogeneous storage cases, Theorem~\ref{Thm2} in fact implies that the optimal download cost in \eqref{general_achievable} depends only on the sum storage space $\sum_{n=1}^{N} m_n$. Thus, any distribution of storage space within the given sum storage space yields the same PIR capacity. In particular, a uniform distribution (the corresponding homogeneous case) has the same PIR capacity. Hence, there is no loss in the PIR capacity due to heterogeneity of storage spaces of the databases.
\end{remark}

\section{Representative Example: $N=3$} \label{capacity}
We introduce the main ingredients of the achievability and converse proofs using the example of $N=3$ databases. Without loss of generality, we take $K=3$ in this section.

\subsection{Converse Proof}
We note that \cite[Theorem~1]{StorageConstrainedPIR} can be applied to any storage constrained PIR problem with arbitrary storage $Z_{1:N}$. Hence, specializing to the case of $N=3$ (and $K=3$) with i.i.d.~messages and uncoded content leads to \cite[eqn.~(39)]{StorageConstrainedPIR},
\begin{align}\label{lb0}
D \geq & L+\frac{4}{27} \sum_{k=1}^{3} H(W_k)+\frac{11}{108} \sum_{i=1}^{3} \sum_{k=1}^{3} H(W_k|Z_i)+\frac{17}{54}\sum_{i=1}^{3} \sum_{k=1}^{3} H(W_k|\mathbf{Z}_{[3]\setminus i})+o(L)
\end{align}
Using the uncoded storage assumption in \eqref{uncoded}, we can further lower bound \eqref{lb0} as,
\begin{align}
D \geq&  L+\frac{4}{27} \sum_{\substack{\cs \subseteq [1:3] \\ |\cs| \geq 1}} \sum_{k=1}^{3} |W_{k,\cs}|L+ \frac{11}{108}  \sum_{i=1}^{3} \sum_{\substack{\cs \subseteq [1:3]\setminus i \\ |\cs| \geq 1}} \sum_{k=1}^{3} |W_{k,\cs}|L\notag\\
&+\frac{17}{54}\sum_{i=1}^{3} \sum_{k=1}^{3} |W_{k,\{i\}}|L+o(L) \\
=&L + \frac{2}{3} \sum_{\substack{\cs \subseteq [1:3] \\ |\cs|= 1}} \sum_{k=1}^{3} |W_{k,\cs}|L+\frac{1}{4} \sum_{\substack{\cs \subseteq [1:3] \\ |\cs| =2}} \sum_{k=1}^{3} |W_{k,\cs}|L\notag\\
&+\frac{4}{27}\sum_{\substack{\cs \subseteq [1:3] \\ |\cs| =3}} \sum_{k=1}^{3} |W_{k,\cs}|L+o(L)
\end{align}
Normalizing with $L$, taking the limit $L \rightarrow \infty$, and using the definition $\alpha_\cs=\frac{1}{K} \sum_{k=1}^{K}|W_{k,\cs}|$ lead to the following lower bound on the normalized download cost $D^*$,
\begin{align}
D^* \geq & 1+2\sum_{\substack{\cs \subseteq [3] \\ |\cs|= 1}} \alpha_\cs+\frac{3}{4}\sum_{\substack{\cs \subseteq [3] \\ |\cs|= 2}} \alpha_\cs+\frac{4}{9}\sum_{\substack{\cs \subseteq [3] \\ |\cs|= 3}} \alpha_\cs\\
=&3\sum_{\substack{\cs \subseteq [3] \\ |\cs|= 1}} \alpha_\cs+\frac{7}{4}\sum_{\substack{\cs \subseteq [3] \\ |\cs|= 2}} \alpha_\cs+\frac{13}{9}\sum_{\substack{\cs \subseteq [3] \\ |\cs|= 3}} \alpha_\cs \label{lb1}
\end{align}
where \eqref{lb1} follows from the message size constraint \eqref{msg_size}.

We further lower bound \eqref{lb1} by minimizing the right hand side with respect to $\{\alpha_\cs\}_{\cs \subseteq [3]}$ under storage constraints. Thus, the solution of the following linear program serves as a lower bound (converse) for the normalized download cost,
\begin{align}\label{final_lb}
\min_{\alpha_\cs \geq 0} \quad &3(\alpha_1+\alpha_2+\alpha_3)+\frac{7}{4}(\alpha_{12}+\alpha_{13}+\alpha_{23})+\frac{13}{9} \alpha_{123} \notag\\
\st  \quad &\alpha_1+\alpha_2+\alpha_3+\alpha_{12}+\alpha_{13}+\alpha_{23}+\alpha_{123}=1 \notag\\
&\alpha_1+\alpha_{12}+\alpha_{13}+\alpha_{123} \leq m_1\notag\\
&\alpha_2+\alpha_{12}+\alpha_{23}+\alpha_{123} \leq m_2\notag\\
&\alpha_3+\alpha_{13}+\alpha_{23}+\alpha_{123} \leq m_3
\end{align}
where variables $\{\alpha_\cs\}_{|\cs|=1}$ are $\{\alpha_1, \alpha_2, \alpha_3\}$, which represent the content stored in databases 1, 2 and 3 exclusively; variables $\{\alpha_\cs\}_{|\cs|=2}$ are $\{\alpha_{12}, \alpha_{13}, \alpha_{23}\}$, which represent the content stored in databases 1 and 2, 1 and 3, and 2 and 3, respectively; and variable $\{\alpha_\cs\}_{|\cs|=3}$ is $\{\alpha_{123}\}$, which represents the content stored in all three databases simultaneously.

Next, we show that the lower bound expressed as a linear program in (\ref{final_lb}) can be achieved.

\subsection{Achievability Proof}
In the placement phase, let $|W_{k,\cs}|=\alpha_\cs$ for all $k \in [K]$. Assign the partition $W_{k,\cs}$ to the set $\cs$ of the databases for all $k \in [K]$. To retrieve $W_\theta$ privately, $\theta \in [K]$, the user applies the Sun-Jafar scheme \cite{JafarPIR} over the partitions of the files.

The partitions $W_{k,1}$, $W_{k,2}$, $W_{k,3}$ are placed in a single database each. Thus, we apply \cite{JafarPIR} with $N=1$, and download
\begin{align} \label{ach-singles}
K(|W_{k,1}|+|W_{k,2}|+|W_{k,3}|)L=3(\alpha_1+\alpha_2+\alpha_3)L
\end{align}
The partitions $W_{k,12}$, $W_{k, 13}$, $W_{k,23}$ are placed in two databases each. Thus, we apply \cite{JafarPIR} with $N=2$, and download
\begin{align} \label{ach-doubles}
\left(1+\frac{1}{2}+\frac{1}{2^2}\right)(|W_{k,12}|&+|W_{k,13}|+|W_{k,23}|)L=\frac{7}{4}(\alpha_{12}+\alpha_{13}+\alpha_{23})L
\end{align}
Finally, the partition $W_{k,123}$ is placed in all three databases. Thus, we apply \cite{JafarPIR} with $N=3$, and download
\begin{align} \label{ach-triples}
\left(1+\frac{1}{3}+\frac{1}{3^2}\right)|W_{k,123}|L=\frac{13}{9}\alpha_{123}L
\end{align}

Concatenating the downloads, file $W_\theta$ is reliably decodable. Hence, by summing up the download costs in (\ref{ach-singles}), (\ref{ach-doubles}) and (\ref{ach-triples}), we have the following normalized download cost,
\begin{align}
\frac{D}{L}=3(\alpha_1+\alpha_2+\alpha_3)+\frac{7}{4}(\alpha_{12}+\alpha_{13}+\alpha_{23})+\frac{13}{9} \alpha_{123}
\end{align}
which matches the lower bound in \eqref{final_lb} and is subject to the same constraints. Hence, the solution to the linear program in \eqref{final_lb} is achievable, and gives the \emph{exact PIR capacity}.

\subsection{Explicit Storage Assignment} \label{explicit-N-3}
In this section, we solve the linear program in \eqref{final_lb} to find the optimal storage assignment explicitly for $N=3$. To that end, we denote $\beta_\ell=\sum_{\cs:|\cs|= \ell} \alpha_\cs$, i.e.,
\begin{align}
\beta_1&=\alpha_1+\alpha_2+\alpha_3 \label{beta1-def} \\
\beta_2&=\alpha_{12}+\alpha_{13}+\alpha_{23} \label{beta2-def}\\
\beta_3&=\alpha_{123} \label{beta3-def}
\end{align}

We first construct a \emph{relaxed} optimization problem by summing up the three individual storage constraints in \eqref{final_lb} into a single constraint. The relaxed problem is,
\begin{align} \label{relaxed1}
\min_{\beta_i \geq 0} \quad &3\beta_1+\frac{7}{4} \beta_2+\frac{13}{9} \beta_3 \notag\\
\st \quad & \beta_1+\beta_2+\beta_3=1 \notag\\
& \beta_1+2\beta_2+3\beta_3 \leq m_s
\end{align}
where we define the sum storage space $m_s=m_1+m_2+m_3$. Plugging $\beta_1=1-\beta_2-\beta_3$,
\begin{align}\label{relaxed}
\min_{\beta_2,\beta_3 \geq 0} \quad &3-\frac{5}{4} \beta_2-\frac{14}{9} \beta_3 \notag\\
\st \quad & \beta_2+\beta_3 \leq 1 \notag\\
& \beta_2+2\beta_3 \leq m_s-1
\end{align}

Since \eqref{relaxed} is a linear program, the solution lies at the boundary of the feasible set. We have three cases depending on the sum storage space $m_s$.

\paragraph{Regime~1:} When $m_s<1$: In this case, the second constraint in (\ref{relaxed}) requires $\beta_2+2\beta_3 <0$, while we must have $\beta_2,\beta_3 \geq 0$. Hence, there is no feasible solution for the relaxed problem and thus the original problem \eqref{final_lb} is infeasible as well.

\paragraph{Regime~2:} When $1 \leq m_s \leq 2$: In this case, the constraint $\beta_2+\beta_3 \leq 1$ is not binding. Hence, the solution satisfies the second constraint with equality, $\beta_2+2\beta_3=m_s-1$, which is non-negative in this regime. Thus, (\ref{relaxed})  can be written in an unconstrained manner as,
\begin{align}\label{unconstrained}
\min_{\beta_3 \geq 0} \:3-\frac{5}{4}(m_s-1-2\beta_3)+\frac{14}{9}\beta_3=\min_{\beta_3 \geq 0} \: \frac{17}{4}-\frac{5}{4}m_s+\frac{17}{18}\beta_3
\end{align}
The optimal solution for \eqref{unconstrained} is $\beta_3^*=0$ and therefore $\beta_2^*=m_s-1$. From the equality constraint $\beta_1+\beta_2+\beta_3=1$, we have $\beta_1^*=2-m_s$. Next, we map the solution of the relaxed problem in (\ref{relaxed}) to a feasible solution in the original problem in \eqref{final_lb}. From (\ref{beta3-def}), $a_{123}^*=\beta_3^*=0$. Thus, at the boundary of the inequality set of \eqref{final_lb}, we have,
\begin{align}
&\alpha_1+\beta_2-\alpha_{23}=m_1 ~~ \Rightarrow ~~ \alpha_1+m_s-1-\alpha_{23}=m_1 ~~ \Rightarrow ~~ \alpha_1-\alpha_{23}=1-(m_2+m_3) \label{assign1}\\
&\alpha_2+\beta_2-\alpha_{13}=m_2 ~~ \Rightarrow ~~ \alpha_2+m_s-1-\alpha_{13}=m_2 ~~ \Rightarrow ~~ \alpha_2-\alpha_{13}=1-(m_1+m_3) \label{assign2}\\
&\alpha_3+\beta_2-\alpha_{12}=m_3 ~~ \Rightarrow ~~ \alpha_3+m_s-1-\alpha_{12}=m_3 ~~ \Rightarrow ~~ \alpha_3-\alpha_{12}=1-(m_1+m_2)\label{assign3}
\end{align}
Depending on the sign of $1-(m_j+m_k)$, where $j,k \in \{1,2,3\}$, we have different content assignments. The common structure of \eqref{assign1}-\eqref{assign3} is $\alpha_i-\alpha_{jk}=1-(m_j+m_k)$. We assign $\alpha_i=\alpha_{jk}+1-(m_j+m_k)$ if $m_j+m_k \leq 1$ and $\alpha_{jk}=\alpha_i-1+(m_j+m_k)$ otherwise. This ensures that $\alpha_\cs \geq 0$ for all $\cs \subseteq [1:3]$. Using these assignments, we have sub-cases depending on the sign of $1-(m_j+m_k)$. We summarize explicit content assignment for these cases in Table~\ref{assign}, where we take $m_1\geq m_2\geq m_3$ without loss of generality, to reduce the number of cases to enumerate. With these solutions, the optimal normalized download cost in this regime is,
\begin{align}\label{DC1}
D^*=\frac{17}{4}-\frac{5}{4}m_s=\frac{17-15\mu}{4}
\end{align}
where $\mu=\frac{m_1+m_2+m_3}{3}=\frac{m_s}{3}$ corresponds to the average storage size.

\begin{table}[t]
	\centering
	\begin{tabular}{|l|l|}
		\hline
		\quad Case  & \qquad Assignment                                                                                                                                                                                      \\ \hline
		\begin{tabular}[c]{@{}l@{}}$1 \leq m_s \leq 2$ \\$m_1+m_2 \geq 1$\\ $m_1+m_3 \geq 1$\\ $m_2+m_3 \geq 1$\end{tabular} & \begin{tabular}[c]{@{}l@{}}$\alpha_1=2-m_s$\\ $\alpha_2=\alpha_3=0$\\ $\alpha_{12}=m_1+m_2-1$\\ $\alpha_{13}=m_1+m_3-1$\\ $\alpha_{23}=1-m_1$\\ $\alpha_{123}=0$\end{tabular}                   \\ \hline
		\begin{tabular}[c]{@{}l@{}}$1 \leq m_s \leq 2$ \\ $m_1+m_2 \geq 1$\\ $m_1+m_3 \geq 1$\\ $m_2+m_3 \leq 1$\end{tabular} & \begin{tabular}[c]{@{}l@{}}$\alpha_1=2-m_s$\\ $\alpha_2=\alpha_3=0$\\ $\alpha_{12}=m_1+m_2-1$\\ $\alpha_{13}=m_1+m_3-1$\\ $\alpha_{23}=1-m_1$\\ $\alpha_{123}=0$\end{tabular} \\ \hline
		\begin{tabular}[c]{@{}l@{}}$1 \leq m_s \leq 2$ \\ $m_1+m_2 \geq 1$\\ $m_1+m_3 \leq 1$\\ $m_2+m_3 \leq 1$\end{tabular} & \begin{tabular}[c]{@{}l@{}}$\alpha_1=1-(m_2+m_3)$\\ $\alpha_2=1-(m_1+m_3)$\\ $\alpha_3=m_3$\\ $\alpha_{12}=m_s-1$\\ $\alpha_{13}=\alpha_{23}=0$\\ $\alpha_{123}=0$\end{tabular}                 \\ \hline
		\begin{tabular}[c]{@{}l@{}}$1 \leq m_s \leq 2$ \\ $m_1+m_2 \leq 1$\\ $m_1+m_3 \leq 1$\\ $m_2+m_3 \leq 1$\end{tabular} & \begin{tabular}[c]{@{}l@{}}$\alpha_1=1-(m_2+m_3)$\\ $\alpha_2=1-(m_1+m_3)$\\ $\alpha_3=m_3$\\ $\alpha_{12}=m_s-1$\\ $\alpha_{13}=\alpha_{23}=0$\\ $\alpha_{123}=0$\end{tabular}                 \\ \hline
		\begin{tabular}[c]{@{}l@{}}$2 \leq m_s \leq 3$\end{tabular} & \begin{tabular}[c]{@{}l@{}} $\alpha_1=\alpha_2=\alpha_3=0$\\ $\alpha_{12}=1-m_3$\\ $\alpha_{13}=1-m_2$\\ $\alpha_{23}=1-m_1$\\ $\alpha_{123}=m_s-2$\end{tabular}                 \\ \hline
	\end{tabular}
\caption{Explicit content assignment for $N=3$ ($m_1 \geq m_2 \geq m_3$ without loss of generality).}
	\label{assign}
\end{table}

\paragraph{Regime~3:} When $2 \leq m_s \leq 3$: In this case, the solution of \eqref{relaxed} is at the intersection of the constraints $\beta_2+\beta_3 = 1$ and $\beta_2+2\beta_3 = m_s-1$. Hence, we have $\beta_2^*=3-m_s$ and $\beta_3^*=m_s-2$, which are both non-negative in this regime. From the equality constraint $\beta_1+\beta_2+\beta_3=1$, we have $\beta_1^*=0$. Next, we map the solution of the relaxed problem in (\ref{relaxed}) to a feasible solution in the original problem in (\ref{final_lb}). From (\ref{beta1-def}), $\beta_1^*=0$  implies $\alpha_1^*=\alpha_2^*=\alpha_3^*=0$. From (\ref{beta3-def}), $\beta_3^*=m_s-2$ implies $\alpha_{123}^*=m_s-2$. At the boundary of the feasible set of \eqref{final_lb}, we have,
\begin{align}
&\alpha_1+\alpha_{12}+\alpha_{13}+\alpha_{123} = m_1 ~~ \Rightarrow ~~ \alpha_1-\alpha_{23}+\beta_2+\beta_3=m_1\\
&\alpha_2+\alpha_{12}+\alpha_{23}+\alpha_{123} = m_2 ~~ \Rightarrow ~~ \alpha_2-\alpha_{13}+\beta_2+\beta_3=m_2\\
&\alpha_3+\alpha_{13}+\alpha_{23}+\alpha_{123} = m_3 ~~ \Rightarrow ~~ \alpha_3-\alpha_{12}+\beta_2+\beta_3=m_3
\end{align}
Plugging $\beta_2^*+\beta_3^*=1$ and $\alpha_i^*=0$ for $i \in \{1, 2, 3\}$ leads to the following content assignment,
\begin{align}
\alpha_{23}^*=1-m_1, \quad \alpha_{13}^*=1-m_2, \quad \alpha_{12}^*=1-m_3
\end{align}
With these solutions, the optimal normalized download cost in this regime is,
\begin{align}\label{DC2}
D^*=3-\frac{5}{4}\beta_2-\frac{14}{9}\beta_3=\frac{85}{36}-\frac{11}{36}m_s=\frac{85-33\mu}{36}
\end{align}
This solution is also shown in Table~\ref{assign}.

\section{Optimal Download Cost for the General Problem} \label{general-capacity}
In this section, we give the proof of Theorem~\ref{Thm1}, i.e., show the achievability and the converse proofs for the PIR problem with heterogeneous databases, for general $N$, $K$, $\bm{m}$.

\subsection{General Achievability Proof}\label{achievability}
In this section, we show the achievability for general $N$ databases and $K$ messages. Let $\tilde{D}_\ell$ denote the optimal normalized download cost for the PIR problem with $\ell$ replicated databases \cite{JafarPIR} storing the same $K$ messages, which is achieved using Sun-Jafar scheme \cite{JafarPIR},
\begin{align}
\tilde{D}_\ell=1+\frac{1}{\ell}+\cdots+\frac{1}{\ell^{K-1}} \label{def-D-tilde-ell}
\end{align}

We partition the messages over all subsets of $[1:N]$, such that $|W_{k,\cs}|=\alpha_\cs$ for all $k \in [1:K]$. Using this partitioning, the subsets $\cs$ such that $|\cs|=1$ correspond to a PIR problem with 1 database and $K$ messages. Hence, by applying the trivial scheme of downloading all these partitions, we download $\tilde{D}_1|W_{k,\cs}|L=K\alpha_\cs L$ bits. For the subsets $\cs$ such that $|\cs|=2$, we have a PIR problem with $2$ databases and $K$ messages. Therefore, by applying Sun-Jafar scheme \cite{JafarPIR}, we download $\tilde{D}_2|W_{k,\cs}|L=(1+\frac{1}{2}+\cdots+\frac{1}{2^{K-1}})\alpha_\cs L$ bits, and so on. This results in total normalized download cost of $\sum_{\ell=1}^{N} \sum_{\cs: |\cs|=\ell} \alpha_\cs \tilde{D}_\ell$. The optimal content assignment is obtained by optimizing over $\{\alpha_\cs\}_{\cs: |\cs|\geq 1}$ subject to the message size constraint \eqref{msg_size}, and the individual storage constraints \eqref{individual}. Thus, the achievable normalized download can be written as the following linear program,
\begin{align}
\min_{\alpha_\cs \geq 0} \quad &\sum_{\ell=1}^{N} \sum_{\cs: |\cs|=\ell} \alpha_\cs \left(1+\frac{1}{\ell}+\cdots+\frac{1}{\ell^{K-1}}\right)\notag\\
\st  \quad &\sum_{\cs:|\cs| \geq 1} \alpha_\cs=1 \notag\\
&\sum_{\cs: n \in \cs} \alpha_\cs \leq m_n, \quad n \in [N]
\end{align}
where $\cs \in \mathcal{P}([1:N])$.

\subsection{General Converse Proof}\label{converse}
In this section, we show the converse for general $N$ databases and $K$ messages. The result in \cite[Theorem~1]{StorageConstrainedPIR} gives a general lower bound for a PIR system with $N$ databases and $K$ messages and arbitrary storage contents $Z_{1:N}$ as
\begin{align}
D^* \geq& 1+\sum_{n_1=1}^N \frac{\lambda(N-n_1,1)}{n_1}+\sum_{n_1=1}^N\sum_{n_2=n_1}^N \frac{\lambda(N-n_1,2)}{n_1n_2}\notag\\
&+\cdots+\sum_{n_1=1}^N\cdots\sum_{n_{K-1}=n_{K-2}}^{N} \frac{\lambda(N-n_1,K-1)}{n_1n_2 \cdots n_{K-1}}
\end{align}
where $\lambda(n,k)$ is given by,
\begin{align}
\frac{1}{KL\binom{K-1}{k}\binom{N}{n}} \sum_{|\mathcal{K}|=k}\sum_{|\mathcal{N}|=n}\sum_{j \in [K]\setminus \mathcal{K}} H(W_j|\mathbf{Z}_\mathcal{N},\mathbf{W}_\mathcal{K})
\end{align}
For uncoded placement, we have,
\begin{align}
H(W_j|\mathbf{Z}_\mathcal{N},\mathbf{W}_\mathcal{K})=H(W_j|\mathbf{Z}_\mathcal{N})=\sum_{\cs:|\cs|\geq 1} |W_{j,\cs}|L
\end{align}

The simplifications in \cite{StorageConstrainedPIR}, which are intended to deal with the nested harmonic sum, can be applied to the heterogeneous storage as well. Thus, the following lower bound in \cite[(77)]{StorageConstrainedPIR} is a valid lower bound for the normalized download cost for the heterogeneous problem,
\begin{align}\label{gen_lb}
D^*\geq 1+\sum_{\ell=1}^{N} \binom{N}{\ell} \left(\tilde{D}_\ell-1\right)x_\ell
\end{align}
where
\begin{align}
x_\ell=\frac{1}{K\binom{N}{\ell}} \sum_{k=1}^{K} \sum_{\cs:|\cs|=\ell}|W_{k,\cs}| \label{x-l-def}
\end{align}
Substituting (\ref{x-l-def}) in \eqref{gen_lb} leads to,
\begin{align}
D^*&\geq 1+\sum_{\ell=1}^{N} \binom{N}{\ell} \left(\tilde{D}_\ell-1\right) \frac{1}{K\binom{N}{\ell}} \sum_{k=1}^{K} \sum_{\cs:|\cs|=\ell}|W_{k,\cs}| \notag\\
&=1+\sum_{\ell=1}^{N} \sum_{\cs:|\cs|=\ell} \left(\tilde{D}_\ell-1\right) \alpha_\cs\\
&=1+\sum_{\ell=1}^{N} \sum_{\cs:|\cs|=\ell} \alpha_\cs \tilde{D}_\ell -\sum_{\ell=1}^{N} \sum_{\cs:|\cs|=\ell} \alpha_\cs \\
&=\sum_{\ell=1}^{N} \sum_{\cs: |\cs|=\ell} \alpha_\cs \left(1+\frac{1}{\ell}+\cdots+\frac{1}{\ell^{K-1}}\right)
\end{align}
where the last step follows from the message size constraint.

This settles Theorem~\ref{Thm1} by having shown that both achievability and converse proofs result in the same linear program which is given in \eqref{general_achievable}.

\section{Equivalence to the Homogeneous Problem} \label{equivalence}
We prove Theorem~\ref{Thm2}, which implies an equivalence between the solution of \eqref{general_achievable} with heterogeneous storage constraints $\bm{m}$ and the solution of \eqref{general_achievable} with homogeneous storage constraint $\mu=\frac{1}{N} \sum_{n=1}^{N} m_n$ for all databases. To that end, let $\beta_n=\sum_{\cs:|\cs|=n} \alpha_\cs$ as before. By adding the individual storage size constraints in \eqref{general_achievable}, we write the following relaxed problem,
\begin{align}\label{general_relaxed}
\min_{\beta_n \geq 0} \quad &\sum_{n=1}^{N} \beta_n \tilde{D}_n \notag\\
\st \quad & \sum_{n=1}^{N} \beta_n=1 \notag\\
\quad &\sum_{n=1}^{N}n \beta_n \leq m_s
\end{align}
where $m_s=\sum_{n=1}^{N} m_n$, as before, is the sum storage space and $\tilde{D}_n$ is defined in (\ref{def-D-tilde-ell}). The solution of the relaxed problem is potentially lower than \eqref{general_achievable}, since the optimal solution of \eqref{general_achievable} is feasible in (\ref{general_relaxed}). Note that the relaxed problem \eqref{general_relaxed} depends only on the sum storage space $m_s$ and the number of databases $N$. Therefore, the corresponding relaxed problem is the same for all distributions of the storage space among databases under the same $m_s$, including the uniform distribution which results in the homogeneous problem. Thus, in order to show the equivalence of the heterogeneous and homogeneous problems, it suffices to prove that the optimal solution of \eqref{general_relaxed} can be mapped back to a feasible solution of \eqref{general_achievable}.

We write the Lagrangian function corresponding to \eqref{general_relaxed} as,
\begin{align}
\mathcal{L}=\sum_{n=1}^{N} \beta_n \tilde{D}_n-\gamma \sum_{n=1}^{N} \beta_n +\lambda \sum_{n=1}^{N}n \beta_n -\sum_{n=1}^{N} \mu_n \beta_n
\end{align}
The optimality conditions are,
\begin{align}\label{KKT}
\tilde{D}_n-\gamma+n\lambda-\mu_n=0, \quad n \in [N]
\end{align}

We have the following structural insights about the relaxed problem. The first lemma states that, in the optimal solution, there are at most two non-zero $\beta$s.

\begin{lemma}\label{lemma1}
	There does not exist a subset $\mathcal{N}$, such that $|\mathcal{N}|\geq 3$ and $\beta_n>0$ for all $n \in \mathcal{N}$.
\end{lemma}
\begin{Proof}
	Assume for sake of contradiction that there exists $\mathcal{N}$ such that $|\mathcal{N}|\geq 3$. Hence, $\mu_n=0$ for all $n \in \mathcal{N}$. From the optimality conditions in \eqref{KKT}, we have,
	\begin{align}
	\gamma=\tilde{D}_n+n\lambda, \quad n \in \mathcal{N}
	\end{align}
	This results in $|\mathcal{N}|$ independent equations in 2 unknowns ($\gamma$ and $\lambda$), which is an inconsistent linear system if $|\mathcal{N}|\geq 3$. Thus, we have a contradiction, and $|\mathcal{N}|$ can be at most 2.
\end{Proof}

The second lemma states that if two $\beta$s are positive, then they must be consecutive.

\begin{lemma}\label{lemma2}
	If $\beta_{n_1}>0$, and $\beta_{n_2}>0$, then $n_2=n_1+1$.
\end{lemma}
\begin{Proof}
	Assume for sake of contradiction that $\beta_{n_1}>0$, $\beta_{n_2}>0$, such that $n_2=n_1+2$, and that  $\beta_{n_0}=0$ where $n_0=n_1+1$. Then, from the optimality conditions, we have,
	\begin{align}
	\tilde{D}_{n_1}-\gamma+n_1 \lambda&=0 \label{l21}\\
	\tilde{D}_{n_0}-\gamma+(n_1+1) \lambda-\mu_{n_0}&=0 \label{l22} \\
	\tilde{D}_{n_2}-\gamma+(n_1+2) \lambda&=0 \label{l23}
	\end{align}
	Solving for $\mu_{n_0}$ leads to,
	\begin{align}
	\mu_{n_0}=\tilde{D}_{n_0}-\frac{1}{2}(\tilde{D}_{n_1}+\tilde{D}_{n_2})
	\end{align}
	Since $D_n$ is convex in $n$, we have $\tilde{D}_{n_0} \leq \frac{1}{2}(\tilde{D}_{n_1}+\tilde{D}_{n_2})$, which implies $\mu_{n_0} \leq 0$, which is impossible since Lagrange multiplier $\mu_{n_0}\geq 0$, and from Lemma~\ref{lemma1}, $\mu_{n_0} \neq 0$. Thus, we have a contradiction, and we cannot have a zero $\beta$ between two non-zero $\beta$s.
\end{Proof}

The third lemma states that having $m_s$ an integer leads to activating a single $\beta$ only.

\begin{lemma}\label{lemma3}
	$\beta_j=1$ and $\beta_n=0$ for all $n \neq j$ if and only if $m_s=j<N$, where $j \in \mathbb{N}$.
\end{lemma}
\begin{Proof}
	From the optimality conditions, we have,
	\begin{align}
	\tilde{D}_j-\gamma+j \lambda&=0 \label{l31}\\
	\tilde{D}_n-\gamma+n \lambda-\mu_n&=0, \quad n \neq j \label{l32}
	\end{align}	
	Substituting $\gamma$ from \eqref{l31} into \eqref{l32} leads to,
	\begin{align}
	(\tilde{D}_n-\tilde{D}_j)+(n-j)\lambda=\mu_n \geq 0 \label{dummy}
	\end{align}
    Since $j<N$, we can choose an $n>j$. Then, (\ref{dummy}) implies,
	\begin{align}
	\lambda \geq \frac{\tilde{D}_j-\tilde{D}_n}{n-j}
	\end{align}	
	Since $\tilde{D}_n$ is monotonically decreasing in $n$, we have $\lambda \geq c >0$ for some positive constant $c=\frac{\tilde{D}_j-\tilde{D}_n}{n-j}$. Since $\lambda>0$, the inequality $\sum_{n=1}^{N}n \beta_n \leq m_s$ must be satisfied with equality. To have a feasible solution for the two equations $\sum_{n=1}^{N} \beta_n=1$ and $\sum_{n=1}^{N}n \beta_n = m_s$, we must have $m_s=j$ and $\beta_j=1$.
\end{Proof}

The fourth lemma gives the solution of the relaxed problem for non-integer $m_s$.
\begin{lemma}\label{lemma4}
	For the relaxed problem \eqref{general_relaxed}, if $j-1 < m_s <j $, then $\beta_{j-1}^*=j-m_s$ and $\beta_j^*=m_s-(j-1)$.
\end{lemma}
\begin{Proof}
	From Lemma~\ref{lemma1}, at most two $\beta$s should be positive. From Lemma~\ref{lemma3}, exactly two $\beta$s should be positive, as $m_s$ is not an integer here. From Lemma~\ref{lemma2}, the positive $\beta$ should be consecutive, and because of continuity, we must have $\beta_{j-1}>0$ and $\beta_j>0$. Thus, on the boundary, we have,
	\begin{align}
	\beta_{j-1}+\beta_j&=1 \\
	(j-1)\beta_{j-1}+j\beta_j&=m_s
	\end{align}
	Solving these equations simultaneously results in $\beta_{j-1}^*=j-m_s$ and $\beta_j^*=m_s-(j-1)$.
\end{Proof}

Thus, Lemmas~\ref{lemma1}-\ref{lemma4} establish the structure of the relaxed problem: First, since $0\leq m_n \leq 1$ for all $n$, we have $0 \leq m_s \leq N$. If $0\leq m_s < 1$, then there is no PIR possible. If $m_s$ is an integer between 1 and $N$, then only one $\beta$ is positive and it is equal to 1. For instance, if $m_s=j$, then $\beta_j=1$. In this case, only one type of $\alpha$ with $j$ subscripts is positive. If $m_s$ is a non-integer  between 1 and $N$, then two $\beta$s are positive. For instance, if $j-1<m_s<j$, then $\beta_{j-1}$ and $\beta_j$ are positive and equal to $j-m_s$ and $m_s+1-j$, respectively. In this case, two types of $\alpha$s with $j-1$ and $j$ subscripts are positive.

Finally, to show the equivalence of the original linear program in \eqref{general_achievable} and the relaxed linear problem in \eqref{general_relaxed}, we need to show that a feasible (non-negative) solution of \eqref{general_achievable} exists for every optimal solution of \eqref{general_relaxed}. That is, the optimal $\beta$s found in solving \eqref{general_relaxed} can be mapped to a set of feasible $\alpha$s in \eqref{general_achievable}. We note that, we have shown this by finding an explicit solution for the case of $N=3$ in Section~\ref{explicit-N-3}. We give an alternative proof for the case of $N=4$ using Farkas' lemma \cite{boyd2004convex} in Appendix~\ref{appendix}. In the following lemma, we give the proof for general $N$ by using the theory of positive linear dependence in \cite{Davis_PosLin}.

\begin{lemma}\label{lemma5}
	There exists a feasible (non-negative) solution of \eqref{general_achievable} corresponding to the optimal solution of the relaxed problem in \eqref{general_relaxed}.
\end{lemma}
\begin{Proof}
Since the inequality in the constraint set of the relaxed problem \eqref{general_relaxed} is satisfied with equality, the $N$ inequalities in the constraint set of the original problem \eqref{general_achievable} should be satisfied with equality as well. We know from Lemmas~\ref{lemma1}-\ref{lemma4} that only two $\beta$s will be positive, therefore, their expressions in terms of the corresponding $\alpha$s will give two more equations. Assuming that $i<m_s<i+1$, we  have $\beta_i^*=i+1-m_s$ and $\beta_{i+1}^*=m_s-i$; $\beta_i$ is a sum of ${N \choose i}$ $\alpha$s and $\beta_{i+1}$ is a sum of ${N \choose i+1}$ $\alpha$s. Thus, we have $(N+2)$ equations in ${N \choose i}+{N \choose i+1}$ variables; and, we need to show that a feasible solution to these linear equations exists.

We denote this linear system of equations as $\bm{A}\bm{\alpha}=\bm{b}$ where $\bm{\alpha}$ is the vector of $\alpha_{\cs}$, i.e., content assignments, and $\bm{b}$ is the vector of $m_i$ and $\beta_i$, i.e., storage constraints and relaxed problem coefficients, i.e.,
\begin{align} \label{general_alpha}
\bm{\alpha}=\begin{bmatrix}
\alpha_{\cs^1_1} & \alpha_{\cs^2_1} & \cdots & \alpha_{\cs^{N \choose i}_1} &
\alpha_{\cs^1_2} & \alpha_{\cs^2_2} & \cdots & \alpha_{\cs^{N \choose i+1}_2} \end{bmatrix}^T
\end{align}
where
\begin{align}
& |\cs^j_1|=i, \quad   j \in \left\{1,2,\cdots,{N \choose i}\right\}\\
& |\cs^j_2|=i+1, \quad j \in \left\{1,2,\cdots,{N \choose i+1}\right\}
\end{align}
and
\begin{align} \label{general_b}
\bm{b}=\begin{bmatrix}
m_1 & m_2 & \cdots & m_N &\beta_{i}&\beta_{i+1}  \end{bmatrix}^T
\end{align}
Now, $\bm{A}$, an $(N+2) \times \left({N \choose i}+{N \choose i+1}\right)$ matrix of zeros and ones, has the following properties:
\begin{enumerate}
\item Every column of the matrix is unique.
\item First ${N \choose i}$ columns have $i$ 1s and $N-i$ 0s in their first $N$ rows. Last two elements of these columns are all 1s and all 0s, respectively.
\item The remaining ${N \choose i+1}$ columns have $i+1$ 1s and $N-i-1$ 0s in their first $N$ rows. Last two elements of these columns are all 0s and all 1s, respectively.
\item First three properties imply that, in the first $N$ rows of the matrix, every permutation of $i$ 1s and $N-i$ 0s exist in the first ${N \choose i}$ columns; and every permutation of $i+1$ 1s and $N-i-1$ 0s exist in the next ${N \choose i+1}$ columns.
\end{enumerate}
			
To clarify the setting with an example, consider $N=4$ and $1<m_s<2$. In this case, we have $\beta_1^*=2-m_s$ and $\beta_2^*=m_s-1$. Corresponding to $\beta_1$, we have ${4 \choose 1}=4$ $\alpha$s, which are $\alpha_1, \alpha_2, \alpha_3, \alpha_4$ which sum to $\beta_1=2-m_s$. Corresponding to $\beta_2$, we have ${4 \choose 2}=6$ $\alpha$s, which are $\alpha_{12}, \alpha_{13}, \alpha_{14}, \alpha_{23}, \alpha_{24}, \alpha_{34}$ which sum to $\beta_2=m_s-1$. Thus, we have the $\bm{\alpha}$ vector:
\begin{align}
\bm{\alpha}=\begin{bmatrix}
\alpha_{1} & \alpha_{2} & \alpha_{3} & \alpha_{4} &
\alpha_{12} & \alpha_{13} & \alpha_{14} & \alpha_{23}& \alpha_{24} & \alpha_{34} \end{bmatrix}^T
\end{align}
the $\bm{b}$ vector:
\begin{align}
\bm{b}=\begin{bmatrix}
m_1 & m_2 & m_3 & m_4 & 2-m_s & m_s-1  \end{bmatrix}^T
\end{align}
and the $\bm{A}$ matrix:
\begin{align}
\bm{A}=\begin{bmatrix}
1 & 0 & 0 & 0 & 1 & 1 & 1 & 0 & 0 & 0 \\
0 & 1 & 0 & 0 & 1 & 0 & 0 & 1 & 1 & 0 \\
0 & 0 & 1 & 0 & 0 & 1 & 0 & 1 & 0 & 1 \\
0 & 0 & 0 & 1 & 0 & 0 & 1 & 0 & 1 & 1 \\
1 & 1 & 1 & 1 & 0 & 0 & 0 & 0 & 0 & 0\\
0 & 0 & 0 & 0 & 1 & 1 & 1 & 1 & 1 & 1
\end{bmatrix}
\end{align}
Note, in the first 4 rows of $\bm{A}$, in the first 4 columns we have all possible vectors with only one 1, and in the remaining 6 columns we have all possible vectors with two 1s.
	
To prove the existence of a feasible solution for $\bm{A}\bm{\alpha}=\bm{b}$, we show that $\bm{b}$ is always a positive linear combination of columns of $\bm{A}$. From the first statement of \cite[Theorem~3.3]{Davis_PosLin}, we note that if we can find a column of $\bm{A}$, for instance $\bm{u}$, such that for all $\bm{v}$ that satisfy $\bm{b}^T\bm{v}>0$, we have $\bm{u}^T\bm{v}>0$; then $\bm{b}$ is a positive linear combination of the columns of $\bm{A}$. Note that, from the last property of $\bm{A}$, if we can find such a column, then we can find an $\cs \subseteq \{1,\cdots,N\}$ that satisfy one of the following inequalities and vice versa:		
\begin{align}
\sum_{j \in \cs, |\cs|=i}{v_j} + v_{N+1} > 0\label{permut1}\\
\sum_{j \in \cs, |\cs|=i+1}{v_j} + v_{N+2} > 0\label{permut2}
\end{align}
where
\begin{align}
\bm{v}=\begin{bmatrix}
v_1 & v_2 & \ldots & v_{N+2}  \end{bmatrix}^T
\end{align}
			
First, we order the variables $v_i$ and $m_i$, $i \in \{1,\cdots,N\}$ among themselves in the decreasing order and we define $m_i'$ and $v_i'$, $i \in \{1,2,\ldots,N\}$ such that,
\begin{align}
v_1' & \geq v_2' \geq \cdots \geq v_N' \\
m_1' & \geq m_2' \geq \cdots \geq m_N'
\end{align}
Then, we have the following series of inequalities for all $\bm{v}$ that satisfy $\bm{b}^T\bm{v}>0$:
\begin{align}
0 &< \sum_{j=1}^{N} m_j v_j + (i+1-m_s) v_{N+1} + (m_s - i) v_{N+2} \label{fun1}\\
&\leq \sum_{j=1}^{N} m_j' v_j' + (i+1-m_s) v_{N+1} + (m_s - i) v_{N+2} \label{fun2}\\
&\leq \sum_{j=1}^{i} v_j' + (m_s - i) v_{i+1}' + (i+1-m_s) v_{N+1} + (m_s - i) v_{N+2} \label{fun3}\\
&\leq \sum_{j=1}^{i} v_j' + \max\{v_{i+1}'+v_{N+2},v_{N+1}\} \label{fun4}
\end{align}
where in \eqref{fun1}, we use Lemma~\ref{lemma4} and insert the values of $\beta_i$ and $\beta_{i+1}$, and in \eqref{fun2} we use the rearrangement inequality \cite{Inequalities}. We have \eqref{fun3} by using the fact that $m_s = \sum_{j=1}^{N}m_j$ is between $i$ and $i+1$, where each $m_j$ is a real number between 0 and 1, and by redistributing the $m_j'$ values where we maximize the ones that are the coefficients of the largest $v_j'$ values. Next, we observe that, $(m_s - i)v_{i+1}' + (i+1-m_s) v_{N+1} + (m_s - i) v_{N+2}$ is the convex combination of $v_{i+1}'+v_{N+2}$ and $v_{N+1}$, which results in \eqref{fun4}.
Hence, we have,
\begin{align}
\sum_{j=1}^{i}v_j' + \max\{v_{i+1}'+v_{N+2},v_{N+1}\}>0 \label{funfinal}
\end{align}
for all $\bm{v}$ that satisfy $\bm{b}^T\bm{v}>0$. Finally, \eqref{funfinal} shows that we can always find $\cs \subseteq \{1,\cdots,N\}$ that satisfies either \eqref{permut1} or \eqref{permut2}, concluding the proof.
\end{Proof}

\section{Conclusions}
We considered a PIR system where a data center places available content into $N$ heterogeneous sized databases, from which a user retrieves a file privately. We determined the exact PIR capacity (i.e., the minimum download cost) under arbitrary storage constraints. By showing the achievability of the solution of a relaxed problem where all available storage space is \emph{pooled} into a sum storage space, by the original problem with individual storage constraints, we showed the equivalence of the heterogeneous PIR capacity to the corresponding homogeneous PIR capacity. Therefore, we showed that there is no loss in PIR capacity due to database storage size heterogeneity, so long as the placement phase is optimized.

\appendix
\section{Alternative Proof for Lemma~\ref{lemma5} for $N=4$} \label{appendix}
Here, we give an alternative proof of Lemma~\ref{lemma5} for $N=4$ using Farkas' lemma. We illustrate the general idea using the example case $1<m_s<2$. Using Lemma~\ref{lemma4}, we have $\beta_1^*=2-m_s$ and $\beta_2^*=m_s-1$. We want to show the existence of $\alpha_i \geq 0$ and $\alpha_{ij} \geq 0$ for all $i,j$ such that,
	\begin{align}
	\alpha_1+\alpha_{12}+\alpha_{13}+\alpha_{14}&=m_1\\
	\alpha_2+\alpha_{12}+\alpha_{23}+\alpha_{24}&=m_2\\
	\alpha_3+\alpha_{13}+\alpha_{23}+\alpha_{34}&=m_3\\
	\alpha_4+\alpha_{14}+\alpha_{24}+\alpha_{34}&=m_1\\
	\alpha_1+\alpha_2+\alpha_3+\alpha_4&=2-m_s\\
	\alpha_{12}+\alpha_{13}+\alpha_{14}+\alpha_{23}+\alpha_{24}+\alpha_{34}&=m_s-1
	\end{align}
	This is a linear system with 10 unknowns and 6 equations in the form of $\tilde{\bm{A}}\bm{\alpha}=\tilde{\bm{b}}$, where $\tilde{\bm{A}}$ is the coefficients matrix. To show the existence of a non-negative solution, we use Farkas' lemma, which states that there exists a non-negative solution $\bm{\alpha}\geq \bm{0}$ that satisfies $\tilde{\bm{A}}\bm{\alpha}=\tilde{\bm{b}}$ if and only if for all $\bm{y}$ for which $\tilde{\bm{A}}^T \bm{y} \geq \bm{0}$, we have $\tilde{\bm{b}}^T \bm{y}\geq 0$. We transform the system of equations into the reduced-echelon form with:
\begin{align}
\tilde{\bm{A}}=\begin{bmatrix}
1 & 0 & 0 & 0 & 0 & 0 & 0 & -1 & -1 & -1 \\
0 & 1 & 0 & 0 & 0 & -1 & -1 & 0 & 0 & -1 \\
0 & 0 & 1 & 0 & -1 & 0 & -1 & 0 & -1 & 0 \\
0 & 0 & 0 & 1 & -1 & -1 & 0 & -1 & 0 & 0 \\
0 & 0 & 0 & 0 & 1 & 1 & 1 & 1 & 1 & 1
\end{bmatrix}
\end{align}
with
\begin{align}
\bm{\alpha}=\begin{bmatrix}
\alpha_1 & \alpha_2 & \alpha_3 & \alpha_4 & \alpha_{12} & \alpha_{13} & \alpha_{14} & \alpha_{23} & \alpha_{24} & \alpha_{34}\end{bmatrix}^T
\end{align}
and
\begin{align}
\tilde{\bm{b}}=\begin{bmatrix}1-m_s+m_1 & 1-m_s+m_2 & 1-m_s+m_3 &
1-m_s+m_4 & m_s-1\end{bmatrix}^T
\end{align}
Hence, for any $\bm{y}$, $\tilde{\bm{A}}^T \bm{y} \geq \bm{0}$ implies,
\begin{align}
	y_1 &\geq 0 \label{y1}\\
	y_2 &\geq 0 \\
	y_3 &\geq 0 \\
	y_4 &\geq 0 \\
	y_5 &\geq y_3+y_4 \\
	y_5 &\geq y_2+y_4 \\
	y_5 &\geq y_2+y_3 \\
	y_5 &\geq y_1+y_4 \\
	y_5 &\geq y_1+y_3 \\
	y_5 &\geq y_1+y_2 \label{y2}
\end{align}
Now, we need to show $\tilde{\bm{b}}^T \bm{y} \geq 0$. We have the following for $\tilde{\bm{b}} \leq \bm{0}$ (the worst case):
\begin{align}
\tilde{\mathbf{b}}^T \mathbf{y}\notag=& (1-m_s+m_1)y_1+(1-m_s+m_2)y_2+(1-m_s+m_3)y_3\notag\\
&+(1-m_s+m_4)y_4+(m_s-1)y_5 \\
\geq & m_1 y_1+m_2 y_2+(1-m_s+m_3)y_3+(1-m_s+m_4)y_4 \label{solution1}\\
\geq & m_1 y_2+m_2 y_2+(1-m_s+m_3)y_3+(1-m_s+m_4)y_4 \label{solution2}\\
\geq & m_1 y_2+m_2 y_2+(1-m_s+m_3)y_2+(1-m_s+m_4)y_2 \label{solution3}\\
=&(2-m_s)y_2\\
\geq &  0
\end{align}
where \eqref{solution3} follows from \eqref{y1}-\eqref{y2} taking into consideration that $1-m_s+m_3 \leq 0$ and $1-m_s+m_4 \leq 0$.

\bibliographystyle{unsrt}
\bibliography{references_new}
\end{document}